# Meteoroid Stream Formation Due to the Extraction of Space Resources from Asteroids

**Logan Fladeland[(1)], Aaron C. Boley[(2)], and Michael Byers[(3)]**

[(1)] UBC, Department of Physics and Astronomy, 6224 Agricultural Rd, Vancouver, BC V6T 1Z1 (Canada), fladelan@phas.ubc.ca
[(2)] UBC, Department of Physics and Astronomy, 6224 Agricultural Rd, Vancouver, BC V6T 1Z1 (Canada), aaron.boley@ubc.ca
[(3)] UBC, Department of Political Science, 1866 Main Mall C425, Vancouver, BC V6T 1Z1 (Canada), michael.byers@ubc.ca

## ABSTRACT

Asteroid mining is not necessarily a distant prospect. Building on the earlier mission Hayabusa, two spacecraft (Hayabusa2 and OSIRIS-REx) have recently rendezvoused with near-Earth asteroids and will return samples to Earth. While there is significant science motivation for these missions, there are also resource interests. Space agencies and commercial entities are particularly interested in ices and water-bearing minerals that could be used to produce rocket fuel in deep space. The internationally coordinated roadmaps of major space agencies depend on utilizing the natural resources of such celestial bodies. Several companies have already created plans for intercepting and extracting water and minerals from near-Earth objects, as even a small asteroid could have high economic worth.

The low surface gravity of asteroids could make the release of mining waste and the subsequent formation of debris streams a consequence of asteroid mining. Proposed strategies that would contain material during extraction could be inefficient or could still require the purposeful jettison of mining waste to avoid the need to manage unwanted mass. Since all early mining targets are expected to be near-Earth asteroids due to their orbital accessibility, these streams could be Earth-crossing and create risks for Earth and lunar satellite populations, as well as humans and equipment on the lunar surface.

Using simulations, we explore the formation of mining debris streams by directly integrating particles released from four select asteroids. Radiation effects are taken into account, and a range of debris sizes are explored. The simulation results are used to investigate the timescales for debris stream formation, the sizes of the streams, and the meteoroid fluxes compared with sporadic meteoroids. We find that for prodigious mining activities resulting in the loss of a few percent of the asteroid's mass or more, it is possible to produce streams that exceed the sporadic flux during stream crossing for some meteoroid sizes.

The results of these simulations are intended to highlight potential unintended consequences that could result from NewSpace activity, including for the future development of international space law. Although the 1967 Outer Space Treaty established that celestial bodies may not be subject to "national appropriation", it did not directly address the extraction of space resources by commercial actors. Based on one of several possible interpretations of the Outer Space Treaty, in 2015 the United States enacted domestic legislation according its citizens the right to possess and sell space resources. Luxembourg established its own domestic legal framework in 2017 and provided financial incentives for space mining companies to incorporate in Luxembourg. Russia has recently announced that it is exploring similar legislation. These three countries might regulate their companies responsibly, but how will the international community respond if space mining companies incorporate in flag-of-convenience states? Will new international rules be adopted to prevent negative human-caused changes to the space environment?

# 1 INTRODUCTION

Asteroid sample return missions such as Hayabusa2 [1] and OSIRIS-REx [2] are mainly motivated by Solar System science. However, in situ resource utilization (ISRU, i.e., space mining and use of those materials) will be relevant in the future and could even become a principal driver for some missions. The International Space Exploration Coordination Group (ISECG) has published "The Global Roadmap for Space Exploration" [3], outlining mutual interests among 14 national space agencies, with ISRU listed as a high priority. This is motivated, in part, by the potential for sourcing fuel in space, as well as manufacturing mission components on another planetary body, which could in turn lower mission costs, enable new types of missions, and facilitate prolonged mission durations. The OSIRIS-REx mission explicitly notes resource prospecting as one of its motivating factors [4].

ISRU is not limited to national space agencies. NewSpace commercial actors also see the potential for resource extraction and related activities (e.g., [5]), with space mining potentially leading to a large off-Earth economy. Although widespread mining of asteroids, the Moon, and other bodies may be decades away, debates over the interpretation of the Outer Space Treaty's prohibition of "national appropriation" are fully engaged (see section 2). Domestic laws have already been enacted in the US and Luxembourg, based on an interpretation that commercial mining does not amount to national appropriation. However, neither international space law nor these new domestic laws are sufficient for addressing the potential complexity of ISRU, including multiple actors, flag of convenience states, and issues of use-versus-appropriation, environmental protection, and safety and sustainability.

For all these reasons, we seek to explore potential consequences of resource extraction from large near-Earth asteroids. While there are many other potential issues that could be addressed (e.g., alteration of pristine cosmochemistry, liability for damage caused by asteroid redirection, data sharing vs. proprietary practices), we choose to focus on one possibility: the alteration of the near-Earth meteoroid environment.

This paper is organized as follows. Additional legal context is given in section 2, followed by simulation and analysis methods in section 3. Section 4 presents the principal results, which are discussed further in section 5. We conclude with section 6, briefly highlighting possible actions that could help to mitigate unintended consequences associated with asteroid-based ISRU.

# 2 LEGAL CONTEXT AND ADDITIONAL MOTIVATION

Despite the technical difficulties ahead, some countries are preparing the legal ground for ISRU. The 1967 Outer Space Treaty (OST) [6] serves as the foundation of international space law, along with the Rescue Agreement, the Liability Convention, and the Registration Convention [7]. Article 1 of the OST states: "Outer space, including the moon and other celestial bodies, shall be free for exploration and use by all States without discrimination of any kind, on a basis of equality and in accordance with international law, and there shall be free access to all areas of celestial bodies." Article 2 states that "Outer space, including the moon and other celestial bodies, is not subject to national appropriation by claim of sovereignty, by means of use or occupation, or by any other means." However, the OST neither defines "national appropriation" nor draws a distinction between claiming ownership of a celestial body and generating revenue from it. Some experts point to Article 31 of the 1969 Vienna Convention on the Law of Treaties ("A treaty shall be interpreted in good faith in accordance with the ordinary meaning to be given to the terms of the treaty in their context and in the light of its object and purpose.") to argue for a broad interpretation of the prohibition on national appropriation, while others draw an analogy with the Law of the Sea, whereby countries and commercial vessels registered by them may engage in fishing on the "high seas" even though this area is, by definition, not subject to national appropriation.

To provide US-based companies with a degree of legal certainty concerning the extraction and use of space resources, the United States Congress adopted the 2015 U.S. Commercial Space Launch Competitiveness Act [8], which accords US citizens the right to own, posses, and sell space resources, at least in the United States. Luxembourg followed suit in 2017, and offered economic incentives for space mining companies to incorporate there [9].

However, these laws focus on ownership issues and do not address safety and sustainability. The Hague International Space Resource Working Group has developed a set of building blocks for updating international space law [10], which could constitute an important step toward addressing safety and sustainability, but these building blocks remain incomplete and do not address the preservation of scientifically valuable materials or other unintended consequences such as debris formation. The work presented here, while dealing with hypothetical situations, nevertheless identifies that new, more comprehensive international rules are necessary.

## 3 METHODS

Our general approach is as follows: we select four different asteroids that have orbital characteristics that are illustrative of those that may be preferable for mining operations. We then integrate each asteroid forward for 100 yr. During the first 10 yr of each simulation, particles are released at a continuous interval, envisaged to be due to mining by either the purposeful jettison of mining waste and/or unintended mass loss due to manipulating an asteroid's surface. Radiation pressure effects are taken into account, assuming six different particle sizes for each asteroid. We do not explicitly fix a set ejection mass fraction for the asteroid (i.e., the mined mass). All particles are effectively massless and are used as tracers of the resulting meteoroid population. Masses are, instead, assigned as a postprocessing step so that different mining scenarios can be explored. As a result of this, perturbations onto the asteroids due to mining are not included in this study.

### 3.1 Integrator and Dynamical Model

Simulations are run with Rebound [11] using the Integrator with Adaptive Time Stepping at 15$^{th}$ order (IAS15) [12]. We also use the ReboundX library packages [13] for including radiation forces and general relativistic (GR) corrections. The GR correction is applied for only the Sun using a modified potential, in accordance with [14], which is fast and recovers the correct precession rates. All objects in the simulation are affected by this modified potential.

The radiation forces on the particles are modelled after [15]:

$$\boldsymbol{a} \approx \beta \frac{GM_\odot}{r^2}[(1 - \hat{\boldsymbol{r}} \cdot \boldsymbol{v}/c)\hat{\boldsymbol{r}} - \boldsymbol{v}/c], \tag{1}$$

where $\boldsymbol{r}$ and $\boldsymbol{v}$ are the particle heliocentric position and velocity vectors, respectively, with radial unit vector $\hat{\boldsymbol{r}}$. The β parameter is the ratio of the radiation pressure force and gravitational force magnitudes, i.e.,

$$\beta \equiv F_{rad}/F_{grav} \approx 5.7 \times 10^{-5} Q_{pr}/\rho s, \tag{2}$$

for particle material density $\rho$ and radius $s$ in cgs units. The radiation coefficient $Q_{pr} \approx 1$. Altogether, the above equations include radiation pressure and Poynting-Robertson drag on the particle dynamics.

The dynamical model for all simulations includes the Sun, Moon, and all eight planets. Asteroids and the meteoroid tracer particles are all treated as massless. For simplicity, the Yarkovsky effect [16] is excluded for these simulations. Initial conditions for all bodies, including the selected asteroids, were obtained using NASA's JPL HORIZONS Web-Interface, with April 13, 2018 00:00:00.0000 TDB (2458221.5 JD) as the starting time[1]. This time was selected, instead of a future time, to avoid implying an anticipated starting date for substantial asteroid-based ISRU.

### 3.2 Debris Ejection

We do not know at this time how asteroid mining will proceed. We take the position here that actors will minimize the mass required for retention and/or transport, which is an independent consideration from how the mining is technically done. We thus assume that all unwanted mass, which accounts for most of the material, is ejected throughout the mining duration. We further note that the mining process could lead to secondary events, such as sporadic mass losses due, for example, to the sublimation of volatiles as new surfaces are exposed to thermal stresses.

To simulate mass loss, we release 3000 particles for each asteroid and each $\beta$ at regular time intervals over a period of 10 yr. We explore $\beta = $ 0.1, 0.03, 0.01, 0.003, 0.001, and 0.0003, which correspond to sizes of 2.2, 7.6, 22, 76, 220, and 760 μm, respectively, assuming a meteoroid material density $\rho = 2500$ kg/m$^3$. Every time a particle is ejected from the asteroid, it is released in a random direction. The initial speed of the particle relative to the asteroid is drawn from a Rayleigh distribution with a mode of 1 m/s, representing a conservative, low-energy expulsion scenario. For comparison, the escape speed from Ryugu is approximately 0.37 m/s. Higher ejection speeds would result in wider debris streams for the low-$\beta$ particles.

---
[1] Orbital data retrieved from https://ssd.jpl.nasa.gov/?horizons on June 2018

### 3.3 Asteroid Selection

We use four asteroids as templates for this study (Table 1). They are chosen because they have favourable orbital characteristics, with semi-major axes between 0.9 and 1.2 au and inclinations less than about 6º. Each asteroid also belongs to the C-complex taxonomy group (e.g., [17]), implying that they may have high water content and would be of potential interest for sourcing fuel. We stress that each asteroid is intended to represent a type of object that might be favoured for ISRU, rather than the detailed evolution of the specific asteroids.

All objects that we select are 400 m in diameter or larger, providing potentially large yields for a long-term mining operation. This is only one possibility. Smaller asteroids could be easier to mine and could utilize enclosures to aid in the resource extraction, but at the expense of a smaller resource yield per trip. Both options could be exploited, although we only focus on large-yield asteroids that have taxonomic data.

Table 1: Asteroid data taken from the JPL Small-Body Database[1], unless otherwise indicated, showing the diameter, mass, semi-major axis, eccentricity, inclination, and Minimum Orbital Intersection Distance (MOID). Uncertainties in the orbital elements are much smaller than the table values and are not shown. For mass estimations, we have assumed a density of 1500 kg/m$^3$, except Ryugu and Bennu, both of which are constrained to 1190±20 kg/m$^3$ and 1190±13 kg/m$^3$, respectively. The values (a) are from [19] and (b) from [20].

| Name | D (m) | Mass (kg) | a (au) | e | i | MOID (au) |
|---|---|---|---|---|---|---|
| 65679 (1989 UQ) | 918±10 | $6.1 \times 10^{11}$ | 0.915 | 0.26 | 1.30 | 0.014 |
| 101955 Bennu (1999 RQ36) | 490.0±0.16 [a] | $7.3 \times 10^{10}$ [a] | 1.126 | 0.20 | 6.03 | 0.0032 |
| 162173 Ryugu (1999 JU3) | 896±4 [b] | $4.5 \times 10^{11}$ [b] | 1.190 | 0.19 | 5.88 | 0.00064 |
| 308635 (2005 YU55) | 400 | $5.0 \times 10^{10}$ | 1.157 | 0.43 | 0.34 | 0.00047 |

Finally, all objects in Table 1 have a Minimum Orbital Intersection Distance (MOID) less than 0.05 au. Due to their sizes, they are also classified as a potentially hazardous asteroids (PHAs). While one might seek to limit any type of mining on a PHA, there are currently no rules that would restrict such actions. It should also be kept in mind that even PHAs can have a negligible impact probability, despite their formal designation. For example, the Double Asteroid Redirection Test (DART) [18] will purposefully perturb the PHA Didymos, with no expected change in the already negligible impact risk. One PHA that does pose a potential risk to Earth is the asteroid Bennu, although we include it due to its accessibility.

### 3.4 Stream Analysis

We evaluate the resulting meteoroid streams in several ways. The most straightforward approach is to look at the time evolution of the 3000 tracer particles for each $\beta$. This allows us to examine how quickly the particles spread and whether any sizes remain confined in longitude after 100 yr. However, this sampling is too sparse for determining meteoroid fluxes at any point in the stream, which would otherwise be dominated by discretization effects. Thus, for evaluating fluxes, we orbit-smooth the streams for all $\beta$ as follows: (1) Each of the 3000 meteoroid tracer particles is taken to represent one possible orbit for debris. (2) Using each tracer, a set of 3000 additional "interpolation" particles are spread out evenly in mean anomaly. This assumes that the meteoroid distribution in each stream has been perfectly sheared out after 100 yr, which is a good approximation for the smallest particles and, as we will see, is acceptable for large particles. (3) The interpolation particles are then used as density and flux estimators for locations that are Earth crossing. The mass flux is set by assuming a total stream mass and evenly distributing that mass over all interpolation particles. Each $\beta$ will thus have 9 million interpolation particles.

Next, we estimate meteoroid flux using two methods. First, we discretize the simulation volume into Cartesian cells. Each interpolation particle has its mass added to the nearest 8 grid cells, with the mass distributed according to a cloud-in-cell interpolation [21]. The momentum is also added into the grid cells using the same scheme. After all particles have been assigned to grid cells, the total mass volume and the meteoroid Cartesian velocities are determined. Finally, we sample Earth's orbit using 10,000 locations in mean anomaly. For every grid cell that Earth intersects and has a non-zero meteoroid density, the magnitude of the flux is determined from

$$\boldsymbol{F}_{met} = \rho_{met}(\boldsymbol{v}_\oplus - \boldsymbol{v}_{\mathbf{met}}). \tag{3}$$

Here, $\rho_{met}$ represents the local spatial mass density of meteoroids and $\boldsymbol{v}_{\mathbf{met}}$ is the corresponding meteoroid velocity. An immediate problem with this method is that the result will depend on the grid cell size. A large grid will decrease the effective density and could make the stream too wide. Too small of a grid cell will produce sharp discretization effects. An acceptable solution is found through iteration for each stream to isolate the smallest cell that still provides a smooth meteoroid distribution. The typical Cartesian grid cell size that we use is $\Delta X, \Delta Y, \Delta Z =$ 0.002, 0.002, 0.0005 au, respectively. The $Z$ direction requires a narrow grid spacing due to the small scale heights of the streams, and even this value remains potentially too large.

To avoid the difficulties using a grid, we use a second method for characterizing the streams, which is an SPH-like density and velocity estimator [22]. Here, for any given point of interest, the meteoroid density at a point $j$ can be approximated by

$$\rho_j = \sum_i^N m \exp(-r_{ij}^2/h_j^2)/(\pi^{3/2} h_j^3), \tag{4}$$

where $m$ is the mass of the interpolation particle, N is the number of interpolation particles, $r_{ij}$ is the distance between the $i^{th}$ interpolation particle and point $j$, and $h_j$ is the smoothing length appropriate for location $j$. The smoothing length is initially set to 400,000 km. However, after an initial density distribution is found, a new scale length is determined using $h_j = (m/\rho_j)^{1/3}$. This is repeated until the smoothing lengths converge to a change smaller than 3%. We avoid oversmoothing by enforcing that $h$ cannot become larger than 400,000 km.

Formally, the sums above are over all particles due to the choice of a Gaussian kernel. However, because we only seek a density and flux estimator, we can truncate the sum to include only particles that are within 0.1 au of the point of interest. The average meteoroid velocity at point $j$ can also be found using

$$\boldsymbol{v}_j = \sum_i^N \frac{m\boldsymbol{v}_i}{\rho_j} \exp(-r_{ij}^2/h_j^2)/(\pi^{3/2} h_j^3), \tag{5}$$

for particle velocity $\boldsymbol{v}_i$. Thus, for any point along an orbit, the meteoroid flux can now be determined using the particle smoothing estimators and Eq. (3). We refer to this as the kernel method.

The grid interpolation tends to give smoother results than the kernel. Regardless, both flux estimations provide comparable results, generally within a factor of two for the peak values.

## 4    RESULTS

All streams shear out over the orbit within the 100 yr timeframe of the simulations. For the high-β particles, this is facilitated quickly by radiation forces. The low-β particles, in contrast, do not immediately shear. Nonetheless, three of the four streams still become distributed over all longitudes after 100 yr. This is enabled by close approaches with

Earth, which perturb the meteoroids. Figure 1 highlights this process for the hypothetical Ryugu stream. After 50 yr, the large debris particles can barely be discerned from the parent asteroid's position, but a close encounter between 50 and 60 yr initiates rapid shearing of the meteoroids. The exception is the hypothetical Bennu stream, which only has full azimuthal spreading for meteoroids with $\beta \geq 0.01$ and partial wrapping for all other meteoroid sizes. Given additional time, we would expect the hypothetical Bennu stream to also behave like the other streams after one of Bennu's expected close encounters with Earth, one of which would occur shortly after the 100 yr timescale that we simulated.

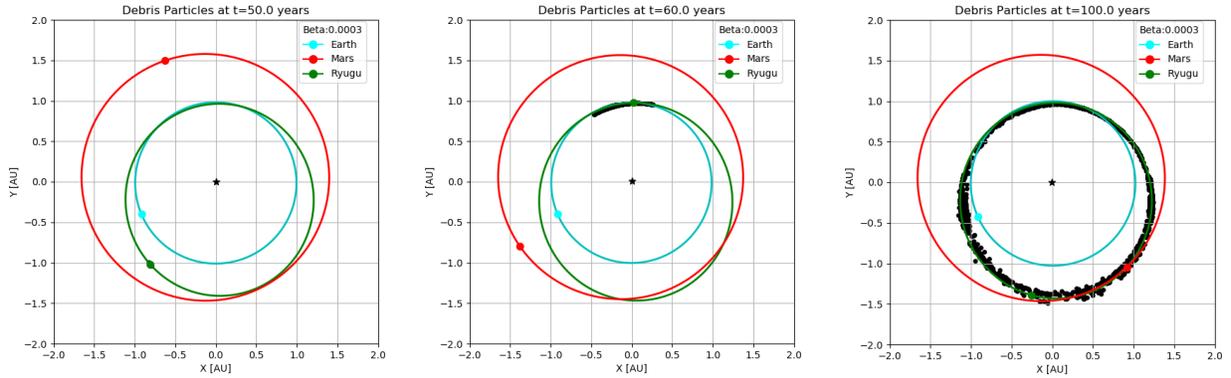

Fig. 1. The *X-Y* scatter plots for the 3000 simulation particles that trace the lowest $\beta$ debris for the hypothetical Ryugu debris stream. For the first 50 yr of the simulation, the particles remain close to Ryugu due to the low $\Delta v$ of the initial mass expulsion. However, a close encounter between 50 and 60 yr led to the rapid spreading of the debris particles, and by 100 yr, the particles were distributed among all longitudes. The positions of Earth, Mars, and Ryugu are shown by the coloured dots. This general behaviour is seen among each of the hypothetical debris streams that we explore, except Bennu, although a similar behaviour is expected after additional integration. The high−$\beta$ particles, in contrast, spread rapidly due to the effects of radiation pressure.

As described in section 3.4, we use orbit-smoothed meteoroid distributions to determine meteoroid fluxes associated with the streams. Fig. 2 shows examples of the resulting meteoroid surface densities in the *X-Y* plane for two particle sizes in the hypothetical Ryugu stream, while Fig. 3 shows all particles for all hypothetical streams. In determining the surface densities, particle masses are interpolated onto a Cartesian grid to obtain volume densities (section 3.4), which are then vertically integrated. We assume that each of the six $\beta$ values contains 1% of the parent asteroid's mass, amounting to 6% of the parent's mass for the entire stream. This high value will be discussed further in section 5, but lower (or higher) values can be inferred by directly scaling the results. Because we do not know what the size distribution of mining debris should be, we use a flat distribution due to its convenience. It can also be re-scaled to explore different scenarios.

For each stream, the peak meteoroid flux (e.g., Fig. 4) is determined and compared with the sporadic meteoroid complex (Fig. 5), assuming 1% of the asteroid's mass goes into each $\beta$. The largest meteoroids that we explore have sufficiently low-$\beta$ that they could, in principle, be used to extend the trend to higher meteoroid masses (assuming a size distribution). Should prodigious resource extraction take place and/or large mass expulsion occur as a result, then human activity could affect the meteoroid environment.

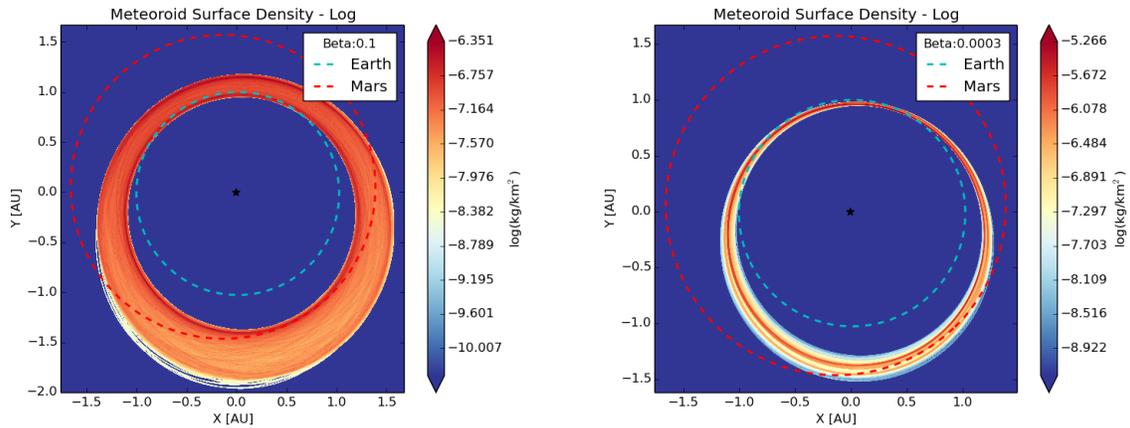

Fig. 2. Surface density plots for the orbit-smoothed debris streams. For setting the mass scale, each $\beta$ value represents 1% of the parent body's original mass. This value can be directly scaled for different mass losses. The orbits are smoothed by using 3000 interpolation particles for every directly simulated particle, as described in section 3.4. This gives an effective resolution of 9 million interpolation particles for generating the map. In the high $\beta$ panel (right), a clear core for the stream can be seen, although there is additional substructure. Note that the colour bar scale is slightly different between panels. The orbits of Earth and Mars are shown for comparison. Recall that this is a surface density plot; Earth crosses this hypothetical stream only once per orbit.

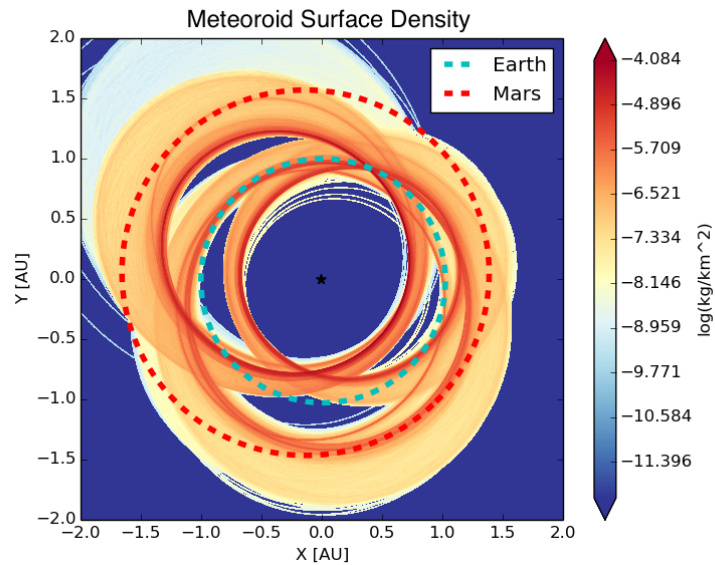

Fig. 3. Similar to Fig. 2, but for all hypothetical streams and particle sizes. The mass density is based on assuming 1% of the asteroid's mass is released in each of the 6 size bins. The resulting stream mass is envisioned to be a combination of the purposeful ejection of mining waste, as well as unintended mass losses (e.g., eruptive events induced by removing surface material). The surface densities can be scaled directly for different mass loss fractions. Earth's and Mars's orbits are shown for comparison.

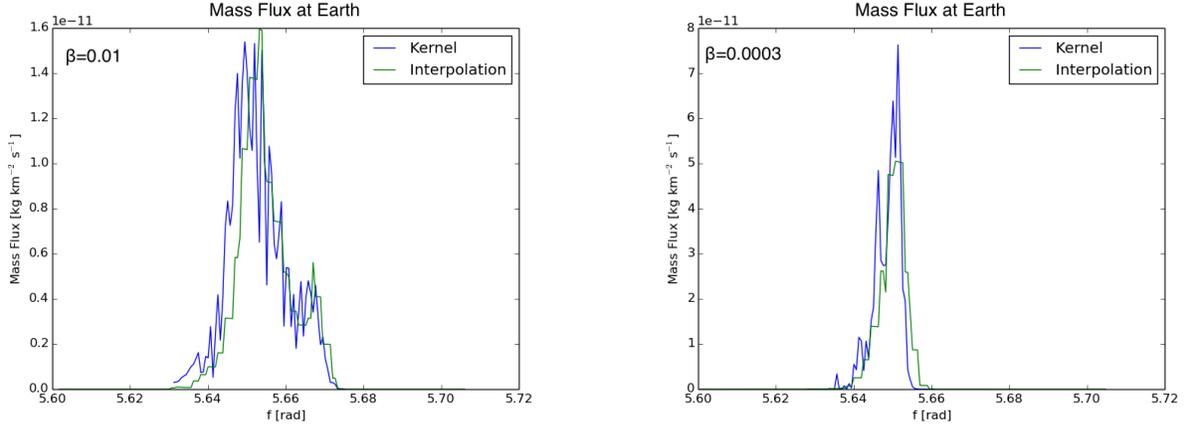

Fig. 4. Mass fluxes as a function of the Earth's true anomaly for the hypothetical Ryugu stream, assuming 1% of the asteroid's mass is ejected for each $\beta$. The mass fluxes can be scaled for different total mass fractions. The blue curves represent the kernel estimator based on the particle distribution, while the green curves correspond to the grid-based interpolation (see section 3.4). The curves give comparable results, although there are differences in the alignment, which are taken to be grid effects. The kernel method is also noisier due to the variable smoothing length. The two $\beta$ values are chosen to highlight the stream variation as a function of meteoroid size.

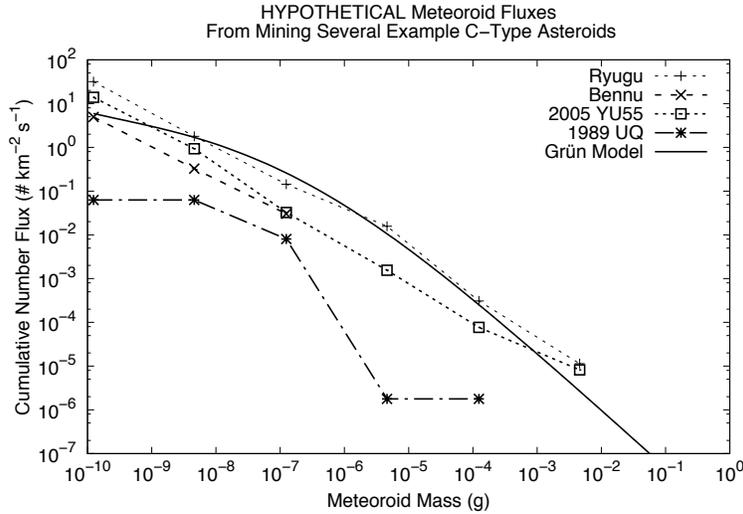

Fig. 5. Cumulative meteoroid flux for the four hypothetical mining streams. Values are based on the peak mass fluxes for each $\beta$, as shown in Fig. 4. We use Eq. 2 to find the particle size for the corresponding $\beta$ and then assume a particle density of 2500 kg/m$^3$ to determine number densities for each bin. For the given mass weighting of each $\beta$, two streams exceed the sporadic meteoroid flux (Grün model [23]) at large meteoroid sizes. If the smallest $\beta$ value (0.0003) is representative of even larger particles, at least for the 100 yr timescale explored here, then this trend could continue depending on the mass distribution. The hypothetical Bennu and 1989 UQ streams do not intersect Earth for the larger meteoroids in these integrations.

## 5 DISCUSSION

Should large NEAs be targeted for resource extraction, then they could, in principle, produce streams with number fluxes that exceed the sporadic meteoroids, although it would require prodigious mass release from the asteroid. We have only considered a flat mass distribution for the amount of material that is ejected at each $\beta$, and have assumed that each bin has 1% of the asteroid's original mass in the stream. For the six particle sizes that we consider, the hypothetical Ryugu stream would thus have a mass of about $2.7 \times 10^{10}$ kg. Over the envisaged 10 yr of mining, this would require $7.4 \times 10^6$ kg per day on average or the consumption of roughly 6200 m³ per day for $\rho = 1200$ kg/m³ (about a soccer pitch that is one metre deep). The total stream mass (6%) would be equivalent to stripping the entire surface of the asteroid to about 18 m deep.

The hypothetical stream from 2005 YU55 is about an order of magnitude less in mass, as well as the corresponding average daily volume consumption (slightly larger than a 25 m x 25 m x 1 m volume). This hypothetical stream's number flux also exceeds the sporadics for the smallest $\beta$, despite the lower mass compared with Ryugu. Such mining would still require multiple machines and significant infrastructure, the feasibility of which is not known. Regardless, the potential for large-scale mining is being explored (e.g., [24,25]). Apart from technical feasibility, there would need to be sufficient demand for ISRU to require such prodigious resource extraction, which will depend strongly on future space traffic, which is also unknown.

In some ways, the results are reassuring, in that the sporadic meteoroid population could be far more significant than any streams produced from asteroid mining, if proper limits are put in place. However, we do not want to dismiss the possibility of secondary effects that could result in large mass expulsions caused by manipulating the asteroid's surface.

These hypothetical streams also have significant differences when compared with some of the major (and real) meteoroid streams [26]. For example, relative speeds tend to only have $v_\infty$ between about 4 and 14 km/s, lower than the major streams, although this is without considering focusing due to Earth.

We have also ignored the possibility of mining small asteroids (in the tens of metre diameter range), which at face value might be more tractable, at least initially. Even if the entire asteroid is effectively reduced to meteoroids, the mass of the stream would be much smaller than that considered here. On the other hand, small asteroids will likely be selected from the population that comes within a few lunar distances of Earth, meaning the resulting streams could be significant despite their mass.

Finally, when releasing particles from the parent body, we only considered an expulsion speed of 1 m/s (for the mode of a Rayleigh distribution). A larger speed would could affect the stream size and decrease the shearing timescale.

# 6 CONCLUSIONS

We have investigated the formation of hypothetical debris streams envisaged to be produced by resource extraction followed by the jettison of mining waste. Four C-Type asteroids were selected based on their accessibility and their potential for large water yields. For the masses that we explored, the meteoroid streams could have peak fluxes that exceed the sporadic population for the smallest meteoroids, although this would require prodigious mining activity with an unknown feasibility. A future that sees extensive human activity in the near-Earth environment may eventually require such resource extraction, which could in turn affect the meteoroid environment. Even in the absence of such largescale mining, it may still be possible to produce the hypothetical streams for even modest resource extraction should mining induce parent body activity followed by natural mass expulsion.

While this may easily be seen as a problem for the future (and from a technical standpoint, it is), the international space law on this issue is contested and several countries have recently enacted domestic legislation that may well shape ISRU. In the absence of widely agreed international rules on the permissibility and conduct of space mining, more and more states can be expected to proceed unilaterally. This situation could lead to the emergence of "flag-of-convenience states" that grant space mining permits with little regard to safety or sustainability and provide little in the way of regulatory oversight. Furthermore, companies may view information concerning their activities on asteroids, as well as composition and telemetry data, as proprietary, preventing the dissemination of that knowledge to the scientific community and preventing the identification of unsafe practices.

Fortunately, it may be possible to establish simple measures that could mitigate some of these concerns, particularly the formation of debris streams with non-trivial mass fluxes. Examples include establishing an international body with the authority to grant mining permits, much like the International Seabed Authority established under the 1982 United Nations Convention on the Law of the Sea. In any scenario, safety and sustainability requirements should be part of the licensing regime. Some of these requirements could limit mining rates or require a company to produce a risk-to-Earth assessment plan. Some asteroids could even be deemed untouchable for safety or scientific reasons. As space law is redefined in the NewSpace era, it must be fully informed by the astrophysical context.

This work was supported in part by The University of British Columbia, an NSERC Discovery Grant, and the Canada Research Chairs program. Computing resources were provided in part by the Canadian Foundation for Innovation and the British Columbia Knowledge Development Fund. AB and MB co-direct the Outer Space Institute, a transdisciplinary institute that addresses challenges associated with space use in the new space era.